\documentstyle[11pt]{article}

\advance \topmargin by -\headheight
\advance \topmargin by -\headsep
     
\evensidemargin \oddsidemargin
\begin{document}
\newcommand{\sect}[1]{\setcounter{equation}{0}\section{#1}}

\topmargin -.6in
\def\nonu{\nonumber}
\def\rf#1{(\ref{eq:#1})}
\def\lab#1{\label{eq:#1}} 
\def\br{\begin{eqnarray}}
\def\er{\end{eqnarray}}
\def\be{\begin{equation}}
\def\ee{\end{equation}}
\def\0{\nonumber}
\def\lb{\lbrack}
\def\rb{\rbrack}
\def\({\left(}
\def\){\right)}
\def\v{\vert}
\def\bv{\bigm\vert}
\def\lskip{\vskip\baselineskip\vskip-\parskip\noindent}
\relax
\newcommand{\nit}{\noindent}
\newcommand{\ct}[1]{\cite{#1}}
\newcommand{\bi}[1]{\bibitem{#1}}
\def\a{\alpha}
\def\b{\beta}
\def\ca{{\cal A}}
\def\cm{{\cal M}}
\def\cn{{\cal N}}
\def\cf{{\cal F}}
\def\d{\delta}
\def\D{\Delta}
\def\eps{\epsilon}
\def\g{\gamma}
\def\G{\Gamma}
\def\grad{\nabla}
\def\h{ {1\over 2}  }
\def\hc{\hat{c}}
\def\hd{\hat{d}}
\def\hg{\hat{g}}
\def\hp{ {+{1\over 2}}  }
\def\hm{ {-{1\over 2}}  }
\def\k{\kappa}
\def\l{\lambda}
\def\L{\Lambda}
\def\lg{\langle}
\def\m{\mu}
\def\n{\nu}
\def\o{\over}
\def\om{\omega}
\def\O{\Omega}
\def\p{\phi}
\def\pa{\partial}
\def\pr{\prime}
\def\ra{\rightarrow}
\def\rh{\rho}
\def\rg{\rangle}
\def\s{\sigma}
\def\t{\tau}
\def\th{\theta}
\def\ti{\tilde}
\def\wti{\widetilde}
\def\inte{\int dx }
\def\xb{\bar{x}}
\def\yb{\bar{y}}

\def\tr{\mathop{\rm tr}}
\def\Tr{\mathop{\rm Tr}}
\def\partder#1#2{{\partial #1\over\partial #2}}
\def\ds{{\cal D}_s}
\def\wtwo{{\wti W}_2}
\def\lie{{\cal G}}
\def\alie{{\widehat \lie}}
\def\dlie{{\cal G}^{\ast}}
\def\elie{{\widetilde \lie}}
\def\edlie{{\elie}^{\ast}}
\def\hlie{{\cal H}}
\def\wlie{{\widetilde \lie}}

\def\rlx{\relax\leavevmode}
\def\inbar{\vrule height1.5ex width.4pt depth0pt}
\def\IZ{\rlx\hbox{\sf Z\kern-.4em Z}}
\def\IR{\rlx\hbox{\rm I\kern-.18em R}}
\def\IC{\rlx\hbox{\,$\inbar\kern-.3em{\rm C}$}}
\def\one{\hbox{{1}\kern-.25em\hbox{l}}}

\def\PRL#1#2#3{{\sl Phys. Rev. Lett.} {\bf#1} (#2) #3}
\def\NPB#1#2#3{{\sl Nucl. Phys.} {\bf B#1} (#2) #3}
\def\NPBFS#1#2#3#4{{\sl Nucl. Phys.} {\bf B#2} [FS#1] (#3) #4}
\def\CMP#1#2#3{{\sl Commun. Math. Phys.} {\bf #1} (#2) #3}
\def\PRD#1#2#3{{\sl Phys. Rev.} {\bf D#1} (#2) #3}
\def\PRB#1#2#3{{\sl Phys. Rev.} {\bf B#1} (#2) #3}

\def\PLA#1#2#3{{\sl Phys. Lett.} {\bf #1A} (#2) #3}
\def\PLB#1#2#3{{\sl Phys. Lett.} {\bf #1B} (#2) #3}
\def\JMP#1#2#3{{\sl J. Math. Phys.} {\bf #1} (#2) #3}
\def\PTP#1#2#3{{\sl Prog. Theor. Phys.} {\bf #1} (#2) #3}
\def\SPTP#1#2#3{{\sl Suppl. Prog. Theor. Phys.} {\bf #1} (#2) #3}
\def\AoP#1#2#3{{\sl Ann. of Phys.} {\bf #1} (#2) #3}
\def\PNAS#1#2#3{{\sl Proc. Natl. Acad. Sci. USA} {\bf #1} (#2) #3}
\def\RMP#1#2#3{{\sl Rev. Mod. Phys.} {\bf #1} (#2) #3}
\def\PR#1#2#3{{\sl Phys. Reports} {\bf #1} (#2) #3}
\def\AoM#1#2#3{{\sl Ann. of Math.} {\bf #1} (#2) #3}
\def\UMN#1#2#3{{\sl Usp. Mat. Nauk} {\bf #1} (#2) #3}
\def\FAP#1#2#3{{\sl Funkt. Anal. Prilozheniya} {\bf #1} (#2) #3}
\def\FAaIA#1#2#3{{\sl Functional Analysis and Its Application} {\bf #1} (#2)
#3}
\def\BAMS#1#2#3{{\sl Bull. Am. Math. Soc.} {\bf #1} (#2) #3}
\def\TAMS#1#2#3{{\sl Trans. Am. Math. Soc.} {\bf #1} (#2) #3}
\def\InvM#1#2#3{{\sl Invent. Math.} {\bf #1} (#2) #3}
\def\LMP#1#2#3{{\sl Letters in Math. Phys.} {\bf #1} (#2) #3}
\def\IJMPA#1#2#3{{\sl Int. J. Mod. Phys.} {\bf A#1} (#2) #3}
\def\AdM#1#2#3{{\sl Advances in Math.} {\bf #1} (#2) #3}
\def\RMaP#1#2#3{{\sl Reports on Math. Phys.} {\bf #1} (#2) #3}
\def\IJM#1#2#3{{\sl Ill. J. Math.} {\bf #1} (#2) #3}
\def\APP#1#2#3{{\sl Acta Phys. Polon.} {\bf #1} (#2) #3}
\def\TMP#1#2#3{{\sl Theor. Mat. Phys.} {\bf #1} (#2) #3}
\def\JPA#1#2#3{{\sl J. Physics} {\bf A#1} (#2) #3}
\def\JSM#1#2#3{{\sl J. Soviet Math.} {\bf #1} (#2) #3}
\def\MPLA#1#2#3{{\sl Mod. Phys. Lett.} {\bf A#1} (#2) #3}
\def\JETP#1#2#3{{\sl Sov. Phys. JETP} {\bf #1} (#2) #3}
\def\JETPL#1#2#3{{\sl  Sov. Phys. JETP Lett.} {\bf #1} (#2) #3}
\def\PHSA#1#2#3{{\sl Physica} {\bf A#1} (#2) #3}
\def\PHSD#1#2#3{{\sl Physica} {\bf D#1} (#2) #3}

\newcommand\twomat[4]{\left(\begin{array}{cc}  
{#1} & {#2} \\ {#3} & {#4} \end{array} \right)}
\newcommand\twocol[2]{\left(\begin{array}{cc}  
{#1} \\ {#2} \end{array} \right)}
\newcommand\twovec[2]{\left(\begin{array}{cc}  
{#1} & {#2} \end{array} \right)}

\newcommand\threemat[9]{\left(\begin{array}{ccc}  
{#1} & {#2} & {#3}\\ {#4} & {#5} & {#6}\\ {#7} & {#8} & {#9} \end{array} \right)}
\newcommand\threecol[3]{\left(\begin{array}{ccc}  
{#1} \\ {#2} \\ {#3}\end{array} \right)}
\newcommand\threevec[3]{\left(\begin{array}{ccc}  
{#1} & {#2} & {#3}\end{array} \right)}

\newcommand\fourcol[4]{\left(\begin{array}{cccc}  
{#1} \\ {#2} \\ {#3} \\ {#4} \end{array} \right)}
\newcommand\fourvec[4]{\left(\begin{array}{cccc}  
{#1} & {#2} & {#3} & {#4} \end{array} \right)}

\begin{titlepage}
\vspace*{-2 cm}
\noindent
\begin{flushright}
\end{flushright}

\vskip 1 cm
\begin{center}
{\Large\bf New Massive Gravity Domain Walls  } \vglue 1  true cm

 { U.Camara dS}$^{*}$\footnote {e-mail: ulyssescamara@gmail.com} and { G.M.Sotkov}$^{*}$\footnote {e-mail: sotkov@cce.ufes.br, gsotkov@yahoo.com.br}\\

\vspace{1 cm}

${}^*\;${\footnotesize Departamento de F\'\i sica - CCE\\
Universidade Federal de Espirito Santo\\
29075-900, Vitoria - ES, Brazil}\\

\vspace{5 cm}

\end{center}

\normalsize
\vskip 0.5cm

\begin{center}
{ {\bf ABSTRACT}}\\
\end{center}

\vspace{.5in}

 The properties of the asymptotic $AdS_3$ space-times representing flat domain walls (DW's) solutions of the New Massive 3D Gravity  with scalar matter are studied. Our analysis is based on  $I^{st}$ order BPS-like equations involving an appropriate superpotential. The Brown-York boundary stress-tensor is used for the calculation of DW's tensions as well as of the $CFT_2$ central charges. The holographic renormalization group flows and the phase transitions  in specific deformed $CFT_2$ dual to  3D massive gravity model with quadratic superpotential are discussed. 
  
\end{titlepage}

\section{Introduction}
The new massive gravity (NMG) represents an appropriate ``higher derivatives'' generalization of 3D Eisntein gravity action:
\begin{eqnarray}
S_{NMG}(g_{\mu\nu},\sigma;\kappa,\Lambda)&=&\frac{1}{\kappa^2}\int dx^3\sqrt{-g}\Big\{\epsilon R +\frac{1}{m^2}{\cal K}-\kappa^2\Big(\frac{1}{2}|\vec{\nabla}\sigma|^2+V(\sigma)\Big)\Big\}\nonumber\\
{\cal K}&=&R_{\mu\nu}R^{\mu\nu}-\frac{3}{8}R^2,\quad \kappa^2=16\pi G,\quad \epsilon=\pm1\label{eq1}
\end{eqnarray}
which unlike its 4D Einstein version is unitary ($i.e.$ ghost-free) under certain restrictions on the matter potential $V(\sigma)$, on the values of the cosmological constant $\Lambda=-\frac{\kappa^2}{2}V(\sigma^*)$ and of the new mass parameter $m^2$ for the both choices of the sign of the $R$-term \cite{1}. It is 1-loop UV finite, but power-counting non-renormalizable \footnote{ although there exist  controversy claims concerning its super-renormalizability \cite{oda}.} quite as in the case of 4D Einstein gravity \cite{flat},\cite{des}. 
The NMG vacuum ($\sigma=const$) sector contains two propagating (massive) degrees of freedom (the ``graviton'' polarizations) and as a result it admits a variety of physically interesting classical solutions - gravitational waves, black holes, etc. \cite{2}. When matter is added the only known exact solutions \cite{3} are certain asymptotically $dS_3$ geometries that describe Bounce-like evolutions of 3D Universe.

The \emph{problem} addressed in the present paper concerns the construction of a family of \emph{flat static domain walls} (DW's) solutions, $i.e.$ $\sigma=\sigma(z)$ and
\begin{eqnarray}
ds^2=dz^2+e^{\varphi(z)}(dx^2-dt^2)\label{eq2}
\end{eqnarray}
of the NMG model (\ref{eq1}) for polynomial matter potentials $V(\sigma)$. We are looking for DW's interpolating between two \emph{different} $AdS_3$ \emph{vacua} $(\sigma_A^*,\Lambda^A_{eff})$, parametrized by the solutions of the algebraic equations:
\begin{eqnarray}
V'(\sigma_{A}^*)=0,\quad\quad \ 2\Lambda_{eff}^A\left(1+\frac{\Lambda^A_{eff}}{4\epsilon m^2}\right) =\epsilon\kappa^2V(\sigma_{A}^*)\label{eq3}
\end{eqnarray}
The study of such DW's is motivated by their important role in the description of the  ``holographic" renormalization group (RG) flows \cite{4} and of the corresponding phase transitions in two-dimensional QFT ``dual'' to  $3D$ massive gravity  (\ref{eq1}). The generalization of the superpotential method proposed in ref. \cite{3} allows an explicit construction of  qualitatively new DW's  relating ``old" to the ``new" purely NMG vacua. Assuming further that the $AdS_3/CFT_2$ correspondence \cite{malda}, \cite{witt},\cite{pol}  takes place for the extended NMG model (\ref{eq1}) as well, we investigate the changes induced by the counter-terms $\cal {K}$ (and by the sign factor $\epsilon$) on the structure of the corresponding $QFT_2$'s $\beta$-function, concerning the  ``$m^2$-corrections" to the  central charges, scaling dimensions and to its free energy. The example of the DW's of  NMG model with \emph{quadratic} superpotential and the phase transitions in its dual perturbed $CFT_2$ ($pCFT_2$) are studied in some details. An extension  of  $d=3$ NMG's BPS-like $I^{st}$ order system  and related to it superpotential to the case of \emph{d-dimensional} New Massive Gravity for $d>3$ is introduced in Sect.7.

\section{Superpotential}
Although the NMG action (\ref{eq1}) involves up to fourth order derivatives of 3D metrics $g_{\mu\nu}$, the corresponding equations for the DW's (\ref{eq2}) are of \emph{second} order:
\begin{eqnarray}
&&\ddot{\sigma}+\dot{\sigma}\dot{\varphi}-V'(\sigma)=0\nonumber\\
&&\ddot{\varphi}\Big(1-\frac{\dot{\varphi}^2}{8\epsilon m^2}\Big)+\frac{1}{2}\dot{\varphi}^2\Big(1-\frac{\dot{\varphi}^2}{16\epsilon m^2}\Big)+\epsilon\kappa^2\Big(\frac{1}{2}\dot{\sigma}^2+V(\sigma)\Big)=0\nonumber\\
&&\dot{\varphi}^2\Big(1-\frac{\dot{\varphi}^2}{16\epsilon m^2}\Big)+\epsilon\kappa^2(-\dot{\sigma}^2+2V(\sigma))=0\label{eq4}
\end{eqnarray}
due to a particular form of the higher derivatives ${\cal K}$-term. A powerful method for construction of analytic non-perturbative solutions of eqs. (\ref{eq4}) consists in the introduction of an auxiliary function $W(\sigma)$ called \emph{superpotential}\footnote{it represents an appropriate $D=3$ NMG adapted version of the Low-Zee superpotential \cite{zee} introduced in the context of DW's solutions of $D=5$ Gauss-Bonnet improuved gravity} \cite{3}, \cite{5} such that 
\begin{eqnarray}
&&\kappa^2V(\sigma)=2(W')^2\Big(1-\frac{\kappa^2W^2}{2\epsilon m^2}\Big)^2-2\epsilon\kappa^2 W^2\Big(1-\frac{\kappa^2 W^2}{4\epsilon m^2}\Big)\nonumber\\
&&\dot{\varphi}=-2\epsilon\kappa W, \quad \quad \dot{\sigma}=\frac{2}{\kappa}W'\Big(1-\frac{\kappa^2W^2}{2\epsilon m^2}\Big)\label{eq5}
\end{eqnarray}
where $W'(\sigma)=\frac{dW}{d\sigma}$, $\dot{\sigma}=\frac{d\sigma}{dz}$ etc. The \emph{statement} is that for each given $W(\sigma)$ all the solutions of the first order system (\ref{eq5}) are solutions of the eqs. (\ref{eq4}) as well. For example, the linear superpotential $W(\sigma)=B\sigma$ ($B=const$) describes a particular double-well matter potential
\begin{eqnarray}
V(\sigma)=\frac{\gamma}{4}\Big(\sigma^2-\frac{m_{\sigma}^{2}}{2\gamma}\Big)^2-\frac{2\Lambda}{\kappa^2}\label{eq6}
\end{eqnarray}   
for $\epsilon m^2>0$ and $\gamma$, $m_{\sigma}^2$ and $\Lambda$ given by
\begin{eqnarray}
\gamma=\frac{2B^4\kappa^2}{m^2}\Big(1+\frac{B^2}{m^2}\Big), \quad m_{\sigma}^2=8\epsilon B^2\Big(1+\frac{B^2}{m^2}\Big), \quad \Lambda=m^2\nonumber
\end{eqnarray} 
The corresponding DW's solutions of eq. (\ref{eq4})
\begin{eqnarray}
\sigma(z)=\frac{\sqrt{2\epsilon m^2}}{B\kappa}\tanh\Big(B^2\sqrt{\frac{2}{\epsilon m^2}}(z-z_{0})\Big),\quad\quad\quad
e^{\varphi(z)+\varphi_{0}}=\Big[\cosh\Big(B^2\sqrt{\frac{2}{\epsilon m^2}}(z-z_{0})\Big)\Big]^{-\frac{2\epsilon|m^2|}{B^2}}\label{eq7}
\end{eqnarray}
have as asymptotics at $z\rightarrow\pm\infty$ two very special NMG - vacua with $\lambda_{BHT}=-\frac{\Lambda}{m^2}=-1$ \cite{1} placed at two degenerate minima
\begin{eqnarray}
\sigma^{\pm}&=&\sigma(z\rightarrow\pm\infty)=\pm\frac{\sqrt{2\epsilon m^2}}{B\kappa}\nonumber
\end{eqnarray}
of the potential (\ref{eq6}) and representing two $AdS_3$ spaces of equal cosmological constant $\Lambda^{\pm}_{eff}=-2\epsilon m^2<0$. 
\section{Vacua and Domain Walls} 
All the  constant $\sigma$  solutions of eqs. (\ref{eq5}) are determined by the real roots of the following algebraic equations:(a) $ W'(\sigma_{a}^*)=0$ and (b) $ W^2(\sigma_{b}^*)=\frac{2\epsilon m^2}{\kappa^2}$,
that describe (a part of) the matter potential $V(\sigma)$ extrema.
Each one of them defines an $AdS_3$ space (i.e. one vacua solution of eqs. (\ref{eq5}))
\begin{eqnarray*}
ds^2=dz^2+e^{-2\epsilon\sqrt{|\Lambda_{eff}^A|}z}(dx^2-dt^2), \quad\quad A=a,b
\end{eqnarray*}
of cosmological constant $\Lambda_{eff}^{A}=-\kappa^2W^2(\sigma_{A}^*)$ as one can see by calculating the values of 3D scalar curvature:
\begin{eqnarray}
R=-2\ddot{\varphi}-\frac{3}{2}\dot{\varphi}^2\equiv8\epsilon(W')^2\left(1-\frac{\kappa^2W^2}{2\epsilon m^2}\right)-6\kappa^2W^2\label{eq8}
\end{eqnarray}
$i.e.$ $R_{vac}=-6\kappa^2W^2(\sigma_{A}^*)=6\Lambda_{eff}^A$. Hence the variety of admissible vacua of NMG model (\ref{eq1}) is defined by the values of the extrema $\sigma^*$ of the matter potential $V(\sigma)$ and by the signs of the parameters $\epsilon$ and $m^2$. For example in the case of \emph{quadratic} superpotential $W_2(\sigma)=B\sigma^2+D$ for  $B>0$ and $D\neq0$ we find one type $(a)$ vacuum $\sigma_a^*=0$ of cosmological constant $\Lambda^{(a)}_{eff}=-\kappa^2D^2$ and for $\epsilon m^2>0$ few type $(b)$ vacua given by
\begin{eqnarray}
\left(\sigma_{\pm}^*\right)^2=\pm\frac{\sqrt{2\epsilon m^2}}{\kappa B}-\frac{D}{B}, \quad \quad \left(\sigma_{-}^*\right)^2\le\left(\sigma_{+}^*\right)^2\label{eq9}
\end{eqnarray}
Depending on the range of values of D ($i.e.$ on the shape of potential $V(\sigma)$) we have:
(1) \emph{no one} type $(b)$ vacuum for $D>\frac{\sqrt{2\epsilon m^2}}{\kappa}$; (2) \emph{two} type $(b)$ vacua $\{\pm |\sigma_{+}^*\}$ for $-\frac{\sqrt{2\epsilon m^2}}{\kappa}<D<\frac{\sqrt{2\epsilon m^2}}{\kappa}$ and (3) \emph{four} type $(b)$ vacua $\{\pm |\sigma_{+}^*,\pm |\sigma_{+}^*|\}$ for  $D<-\frac{\sqrt{2\epsilon m^2}}{\kappa}$.

Note that all the type (b) vacua have by construction \emph{equal} cosmological constants $\Lambda_{eff}^{(b)}=-2\epsilon m^2$.
We consider as an example the DW's one can construct in the \emph{region} (2) above, characterized by the three vacua $\pm |\sigma_+^* |$ and $\sigma_a^*=0$. Then the two DW's solutions of eqs. (\ref{eq5}) connecting $| \sigma_+^*|$ (or $-| \sigma_+^* |$) with $\sigma_a^*$ have the following rather implicit form:
\begin{eqnarray}
&&\left(\sigma^4\right)^{\alpha_{+}\alpha_{-}}\left(\sigma^2+| (\sigma_-^*)^2|\right)^{-\alpha_-}\left((\sigma_+^*)^2-\sigma^2\right)^{-\alpha_+}=e^{\frac{16B}{\kappa}(z-z_0)}\nonumber\\
&&e^{\varphi-\varphi_0}=\left(\sigma^2\right)^{-\frac{D\kappa^2\alpha_+\alpha_-}{4\epsilon B}}
\left((\sigma_+^*)^2-\sigma^2\right)^{\frac{\alpha_+\kappa\sqrt{2\epsilon m^2}}{8\epsilon B}}\left(|(\sigma_-^*)^2|+\sigma^2\right)^{-\frac{\alpha_-\kappa\sqrt{2\epsilon m^2}}{8\epsilon B}}\label{eq10}
\end{eqnarray}
where we have denoted:
\begin{eqnarray*}
\alpha_+=\left(1-\frac{D\kappa}{\sqrt{2\epsilon m^2}}\right)^{-1}, \ \alpha_-=\left(1+\frac{D\kappa}{\sqrt{2\epsilon m^2}}\right)^{-1}.
\end{eqnarray*}
Nevertheless one can easily verify that the corresponding asymptotics (at $z\rightarrow\pm\infty$) of $\sigma(z)$:
\begin{eqnarray}
\sigma(z)\stackrel{z\rightarrow\pm\infty}{\approx}\sigma_{A}^* - \sigma_{A}^{0}e^{\mp2\Delta_A\sqrt{|\Lambda^A_{eff}|}z},\quad\quad
\sigma(\infty)=\pm|\sigma_+^*|, \quad \sigma(-\infty)=\sigma_a^*=0\nonumber\\
\Lambda_{eff}^{\pm}=-2\epsilon m^2,\quad \Lambda_{eff}^a=-\kappa^2D^2,\quad\quad
\Delta_A=1+\sqrt{1-\frac{m_{\sigma}^2(A)}{\Lambda_{eff}^A}},\quad \ m_{\sigma}^2=V''(\sigma_A^*)\label{eq11}
\end{eqnarray}
indeed coincide with the \emph{vacuum data} $(\sigma_A^*,\Lambda_{eff}^A, \Delta_A)$ that determine the boundary conditions for the DW solutions in region (2). Observe that the scale factor $e^{\varphi(z)}$ has \emph{different} asymptotic behaviour depending on the sign of $D$: in the case of negative values, $i.e.$ for $D\in\left(-\frac{\sqrt{2\epsilon m^2}}{\kappa},0\right)$ and $\epsilon=-1$, $m^2<0$, we find that:
\begin{eqnarray}
e^{\varphi}\stackrel{z\rightarrow\infty}{=}e^{2\sqrt{|\Lambda_+ |}z}\rightarrow\infty,\quad\quad
e^{\varphi}\stackrel{z\rightarrow-\infty}{=}e^{-2\sqrt{|\Lambda_a |}z}\rightarrow\infty\label{eq12}
\end{eqnarray}
while for $\epsilon=-1$, $m^2<0$ and  considering  positive values  within the interval $D\in\left(0,\frac{\sqrt{2\epsilon m^2}}{\kappa}\right)$ we have that:
\begin{eqnarray}
e^{\varphi}\stackrel{z\rightarrow\infty}{=}e^{2\sqrt{|\Lambda_+ |}z}\rightarrow\infty,\quad\quad\
e^{\varphi}\stackrel{z\rightarrow-\infty}{=}e^{2\sqrt{|\Lambda_a |}z}\rightarrow0\label{eq13}
\end{eqnarray}    

As it well known the divergences of the scale factor correspond to $AdS_3$ type of boundaries. The regions of vanishing scale factor (which are not curvature singularities) represent null Cauchy \emph{horizons}, where the causal description in  the Poincare patch terminates. Therefore our DW's (\ref{eq10}) define particular asymptotically $AdS_3$ ($(a)AdS_3$) spaces: the one with $D<0$ (see eqs. (\ref{eq12})) has \emph{two different boundaries} and the other one with $D>0$ (see eqs. (\ref{eq13})) has \emph{one boundary} at $z\rightarrow \infty$ and one \emph{null horizon} at $z\rightarrow-\infty$. Let us also mention that the linear superpotential DW's (\ref{eq7}) (and more general all the DW's relating two type (b) vacua) in the case $\epsilon=-1$, $m^2<0$ describe $(a)AdS_3$ spaces of \emph{two boundaries}, but in this case of \emph{equal} cosmological constants $\Lambda_{eff}^{b_{\pm}}=-2\epsilon m^2$.
\section{Unitarity and BF - conditions}
The Bergshoeff-Hohm-Townsend (BHT) unitarity conditions \cite{1}:
\begin{eqnarray}
m^2\left(\Lambda_{eff}^A-2\epsilon m^2\right)>0\label{A1}
\end{eqnarray}
together with the Higuchi bound:
\begin{eqnarray}
\Lambda_{eff}^A\le M_{gr}^2(A)=-\epsilon m^2+\frac{1}{2}\Lambda_{eff}^A\label{A2}
\end{eqnarray}
for the massive spin two field (``graviton'') are result of the requirement of perturbative (1 - loop) unitarity consistency of NMG model (\ref{eq1}) for $\sigma=const$. When massive scalar field is also included one have to further impose the Breitenlohner-Freedman (BF) condition \cite{BF} which for $D=3$ reads:
\begin{eqnarray}
\Lambda_{eff}^A\le m_{\sigma}^2(\sigma_{A}^*)=V''(\sigma_{A}^*)\label{A3}
\end{eqnarray}
or in its \emph{stronger} form:
\begin{eqnarray}
\Lambda_{eff}^A\le m_{\sigma}^2(\sigma_{A}^*)<0\label{A4}
\end{eqnarray}

It is convenient to parametrize the effective ``vacuum masses'' $m_{\sigma}^2(\sigma_A^*)=\kappa^2 W_A^2 y_A (y_A-2)$ in terms of the scaling dimensions $\Delta_A=2-y_A$ of 2D field $\phi_{\sigma}(x,t)$ ``holographically dual'' of $\sigma(z)$ \cite{4} ($A=a,b$):
\begin{eqnarray}
y_{a}=y(\sigma_{a}^*)=\frac{2\epsilon W_a''}{\kappa^2 W_a}\left(1-\frac{\kappa^2 W_a^2}{2\epsilon m^2}\right),\quad\quad\quad
y_b=y(\sigma_b^*)=-\frac{4\epsilon (W_b ')^2}{\kappa^2 W_b^2}\label{A5}
\end{eqnarray}
Let us consider the case $\epsilon=-1$, $m^2<0$. Then the above unitarity conditions (\ref{A1}), (\ref{A2}), (\ref{A4}) take the following simple form
\begin{eqnarray}
0\le \frac{\kappa^2 W_A^2}{2\epsilon m^2}\le2, \quad\quad  0\le y_a<2\label{A6}
\end{eqnarray}
which imposes  restrictions on the values of the parameters of NMG model (\ref{eq1}). For example, in the case of linear superpotential both vacua satisfy all the unitarity conditions  only for specific values of the parameter $B$ such that:  $B^2\le |m^2|$. In the particular case of quadratic superpotential of only three vacua $\pm |\sigma_{+}^*|$ and $\sigma_a^*=0$ ($i.e.$ in region (2)) all the vacua are \emph{unitary} and satisfy the weak BF condition, when the superpotential parameters  are restricted as follows: $\kappa^2 D^2 < 2\epsilon m^2$ and $B>0$.
Therefore the corresponding DW's (\ref{eq10}) are interpolating between two \emph{unitary} vacua. Such DW's turns out to also have positive tensions $\tau_{DW}>0$ as it shown in Sect. 6.
\section{Domain Walls Tensions}
In all the ``planar'' DW's (2) of NMG model (\ref{eq1}) the scalar matter  is uniformly distributed ($i.e. \ \frac{\partial\sigma}{\partial x}=0$) along the whole $x$-axis and therefore such DW's have infinite energy. As it well known \cite{6}, an important characteristics of the gravitational properties of such DW's is given by the values of their \emph{energy densities} $\epsilon_{DW}=\frac{E_{DW}}{L_x}$ (equals of their tensions $\tau_{DW}$). In the case of $(a)AdS_3$ geometries it is given by \cite{7}:
\begin{eqnarray}
\tau_{DW}=\lim_{L_x\rightarrow\infty}\frac{1}{L_x}\sum_{A=\pm}v_A\int_{-L_x/2}^{L_x/2}dx\xi^iT_{ij}^{(A)}\xi^{j},\quad\quad\quad i,j=0,1\label{eq14}
\end{eqnarray}
where $A=\pm$ denote the two $z\rightarrow\pm\infty$ limits $(\partial M)_{A}$ describing $(a)AdS_3$ boundaries or/and horizons; $v_{\pm}=\pm1$ and $\xi^\mu=(0,\xi^i)$ is time-like Killing vector, orthogonal to both $(\partial M)_A$ - surfaces and normalized as $\xi^i\gamma_{ij}^A\xi^i=-1$. The Brown-York ``boundary'' stress-tensor $T_{ij}^{(A)}$ \cite{bala} is defined as follows:
\begin{eqnarray}
T_{ij}^{(A)}=-\frac{2}{\sqrt{-\gamma^A}}\frac{\delta S_{NMG}^{BY}}{\delta \gamma_{A}^{ij}}v_A\label{eq15}
\end{eqnarray}
where $\gamma_{ij}^{A}$ are the corresponding ``boundary/horizon'' $(\partial M)_A$ - metrics:
\begin{eqnarray*}
\gamma_{ij}^A(x,t)=\lim_{z\rightarrow\pm\infty}\gamma_{ij}(x,t|z), \quad\quad \gamma_{ij}(x,t|z)=e^{\varphi(z)}\eta_{ij},\quad\quad\quad
\eta_{ij}=diag(+,-)
\end{eqnarray*}
The main ingredient of the NMG version of the Brown-York formula (\ref{eq14}) and (\ref{eq15}) is the improved NMG action $S_{NMG}^{BY}=S_{NMG}+S_{gGH}$ with few ``boundary'' terms $S_{gGH}$ added. They represent an appropriate generalization of the Gibbons-Hawking boundary action to the case of NMG model (\ref{eq1}) recently proposed by Hohm and Tonni \cite{8}:
\begin{eqnarray}
S_{gGH}=-\frac{2}{\kappa^2}\sum_{A=\pm}v_{A}\int_{(\partial M)_A}dxdt\sqrt{-\gamma}\Big(\epsilon K-\frac{1}{2}fK+\frac{1}{2}f_{ij}K^{ij}\Big)\label{eq16}
\end{eqnarray}
 where $K_{ij}$ is the extrinsic curvature of 2D ``boundary'' surface $(\partial M)_A$; $f_{\mu\nu}$ is the auxiliary Pauli-Fierz spin two field \cite{1} whose ``on-shell'' form
 \begin{eqnarray*}
f_{\mu\nu}=\frac{2}{m^2}\left(R_{\mu\nu}-\frac{1}{4}g_{\mu\nu}R\right), \ \mu;\nu=0,1,2
\end{eqnarray*}
is used in eq. (\ref{eq16}); $f=\gamma^{ij}f_{ij}$ and $K=\gamma^{ij}K_{ij}$. In the case of DW's (2) one can further apply the $I^{st}$ order equations (\ref{eq5}) in order to derive the following simple ``boundary'' form of the improved action \footnote{unlike the case of the Einstein gravity where all the flat DW's are of BPS type \cite{town}, for the 3D NMG  DW's one need to use the $I^{st}$ order eqs.(\ref{eq5}) in order to prove that the remaining terms in the bulk action represent a total derivative.} :
\begin{eqnarray}
S_{NMG}^{BY}(DW)=-\frac{2}{\kappa}\sum_{A=\pm}v_A\int_{(\partial M)_A} dxdt\sqrt{-\gamma} W(\sigma)\left(1+\frac{\kappa^2 W^2(\sigma)}{2\epsilon m^2}\right)\label{eq17}
\end{eqnarray}
Then according to  the definitions (\ref{eq15} ) and (\ref{eq16}) one  easily obtains the corresponding explicit form of the  ``boundary'' stress-tensor for the NMG model with scalar matter:  
\begin{eqnarray}
T_{ij}^A(DW)=-\frac{2}{\kappa}W(\sigma_A^*)\left(1+\frac{\kappa^2W^2(\sigma_A^*)}{2\epsilon m^2}\right)\gamma^A_{ij}\label{eq18}
\end{eqnarray}
which allows us  to calculate the values of the  DW's tensions:
\begin{eqnarray}
\tau_{DW}=\frac{2}{\kappa}\sum_{A=\pm}v_AW_A\left(1+\frac{\kappa^2 W_A^2}{2\epsilon m^2}\right),\quad\quad W_A = W(\sigma_{A}^*) \label{eq19}
\end{eqnarray}
Note that in the $m^2\rightarrow\infty$ limit the above formula reproduces the well know results for a flat DW's tensions in 3D Einstein gravity  obtained by the Israel's thin wall approximation \cite{6}.
\section{Boundary counter-terms and central charges}

Consider NMG model (\ref{eq1}) in the limit of small  effective cosmological constant: $L_A\gg G=l_{pl}$,            where 
$|\Lambda^A_{eff}|L_A^2=1$. The  $AdS_3/CFT_2$ correspondence suggests that each of its vacua $(\sigma_A^*,\Lambda_{eff}^A, \Delta_A)$ determine the main features of certain $CFT_2$ , ``living" on the corresponding
2-D boundaries/horizons $(\partial M)_A$ of $(a)AdS_3$ space-times (\ref{eq11}). As in the case of 3D Einstein Gravity (i.e. the $m^2\rightarrow\infty $ limit of (\ref{eq1})), all the 2D data, namely: central charges, scaling dimensions $\Delta_A(\sigma^*_A)=2-y_A$ and the vacuum expectation values 
\begin{eqnarray}
<A_{vac}|\hat{T}_{ij}^A(x_+,x_-)|A_{vac}>=-\frac{2}{\sqrt{-\gamma^A}}\frac{\delta S_{NMG}^{ren}(DW)}{\delta \gamma_{A}^{ij}(x_+,x_-)}=T_{ij}^A + T^{ct,A}_{ij}= T_{ij}^{ren}(A),\quad x_{\pm}=x\pm t \label{eq20}
\end{eqnarray}
$<A_{vac}|\hat{\Phi}_{\sigma}(x_+,x_-)|A_{vac}>$, etc. of 2D operators $\hat{T^A_{ij}}$ and $\hat{\Phi}_{\sigma}$, duals of the 3D NMG model fields $\gamma_{ij}$ and $\sigma$ - can be extracted from NMG classical action  (\ref{eq17}), appropriately renormalized: $S^{ren}_{NMG}=S^{BY}_{NMG}+S^{ct}_{NMG}$, where
\begin{eqnarray}
S_{NMG}^{ct}=\frac{2}{\kappa}\sum_{A=\pm}v_A\int_{(\partial M)_A} dxdt\sqrt{-\gamma} W(\sigma)\left(1+\frac{\kappa^2 W^2(\sigma)}{2\epsilon m^2}\right)\label{eq21}
\end{eqnarray}
The particular form\footnote{although all our arguments are based on specific DW's solutions of NMG model(\ref{eq1}) it is expected that as in the Einstein gravity case, this form of the counter-terms is universal, i.e. itcancels the $S_{NMG}^{BY}$ divergences for larger class of solutions.} of the boundary counter-terms (\ref{eq21}) we have introduced above is a consequence of the condition that the vacua NMG solutions are conjectured to describe the vacua states $|A_{vac}>=|\sigma^*_A,\Lambda^A_{eff}>$  of   corresponding (UV and IR) $CFT_2$'s, which by definition must have vanishing dimensions $\Delta^A_{vac}=0$ and energy $ E^A_{vac}$=0 (for planar 2D geometries), i.e. 
\begin{eqnarray}
<\hat{T}_{ij}^A(x_+,x_-)>_A = 0 = T_{ij}^A + T^{ct,A}_{ij} \label{eq22}
\end{eqnarray}
Note that $S^{ct}$ makes NMG action convergent, i.e. we have $S^{ren}_{NMG}(DW)=0$, thus providing hints that such DW's are  stable.

    The central  charges $c_A$ of these $CFT_2$'s are given by the normalization constants of the stress-tensor's 2-point functions
\begin{eqnarray}
<\hat{T}_{\pm\pm}^A(x_{\pm})\hat{T}_{\pm\pm}^A(0)>_A = \frac{c_A}{2 x_{\pm}^4} \label{eq23}
\end{eqnarray}
or equivalently by the coefficients of the inhomogeneous part of the stress-tensor's transformation laws:
\begin{eqnarray}
<\delta_{\xi}\hat{T}_{\pm\pm}^A(x_{\pm})>_A =-\frac{c_A}{24\pi}\xi_{\pm}^{'''} \label{eq24}
\end{eqnarray}
under infinitesimal 2D  transformations: $x^{'}_{\pm}= x_{\pm}+\xi_{\pm}(x_{\pm})$. According to the Brown-Henneaux's observation \cite{9} the 3D counterparts of these 2D
conformal symmetries are given by special 3D  diffeomorphisms that keep invariant the asymptotic form of the $AdS_3$ metrics\footnote{but are larger then the $AdS_3$ isometry group $SO(2,2)$} such that 
\begin{eqnarray}
\delta_{\xi}\gamma_{\pm\pm}(x_+,x_-) =-\frac{L_A^2}{2}\xi_{\pm}^{'''}. \label{eq25}
\end{eqnarray}
As a result  the corresponding improved Brown-York stress-tensor $T_{ij}^{ren}(A)$ (proportional to $\gamma_{ij}$) gets inhomogeneous terms under these transformations:
\begin{eqnarray}
\delta_{\xi}{T}_{\pm\pm}^{ren}(A) =\frac{L_A^2}{\kappa}W_A\left(1+\frac{\kappa^2 W_A^2}{2\epsilon m^2}\right)\xi_{\pm}^{'''} \label{eq26}
\end{eqnarray}
which allows to calculate the $CFT_2$'s central charges in therms of the NMG vacuum data:
\begin{eqnarray}
c_A =-\frac{3L_A^2}{2G}\kappa W_A\left(1+\frac{\kappa^2 W_A^2}{2\epsilon m^2}\right) \label{eq27}
\end{eqnarray} 
Consider for example the domain wall solution (\ref{eq10}) for quadratic superpotential under the restriction
$0 <\kappa D < \sqrt{2\epsilon m^2}$ and $B > 0 $ , which for $\epsilon=-1$ and $ m^2 < 0$ interpolates between two vacua, i.e. $AdS_3$'s of effective cosmological constants $\Lambda_+=-2\epsilon m^2$ and $\Lambda_a=-\kappa^2D^2$  as one can see from its asymptotic form (\ref{eq13}). Since in this case  we have $W(\sigma)>0$, the following identification $ \kappa W_A=-\frac{\epsilon}{L_A}$ takes place. As a consequence the corresponding central charges of the two $CFT$'s representing these vacua get the familiar form \cite{1},\cite{8}:  
\begin{eqnarray}
 c_A =\frac{3\epsilon L_A}{2G}\left(1+\frac{L_{gr}^2}{L_A^2}\right),\quad\quad L_{gr}^2=\frac{1}{2\epsilon m^2}\gg l_{pl}^2\label{eq28A}
\end{eqnarray}
 The same central charge formula  turns out to be  valid  in the more general case of DW's for which  the  superpotential   $W(\sigma)$ does not change its sign between the two vacua $\sigma_A^*$ and for the case of  ``non-unitary" vacua with $\epsilon = 1$ and $m^2 >0$ as well.
  It is worthwhile to mention an interesting fact that the DW's tensions (\ref{eq19}) can be rewritten in terms of the central charges as follows:
\begin{eqnarray}
\tau_{DW}(L_+,L_a)= -\frac{1}{12\pi}\left(\frac{c_+}{L_+^2} -\frac{c_a}{L_a^2}\right) \label{eq29A}
\end{eqnarray}   
Observe that the condition of  positive tensions, i.e. $\tau_{DW}(L_+,L_a) >0$  requires  $|\Lambda_+|>|\Lambda_a|$  which is automatically satisfied in the example discussed above. In the case of "unitary" BHT -vacua  $\epsilon=-1$ and $m^2<0$ (i.e. for negative $c_A$'s )  this condition is equivalent to the following restriction on the central charges:
\begin{eqnarray}
\frac{|c_a|}{|c_+|} > \frac{L^2_a}{L^2_+} >1  \label{eq29B}
\end{eqnarray} 
i.e. we have $c_+ > c_a$. It turns out that such ``ordering" of the UV and IR central charges   determines the direction of the RG flow in the dual 2D $pCFT_2$ as we are going to show in the next section.
\section{Comments on holographic RG flows and NMG's extensions }

The off-critical $(a)AdS_3/pCFT_2$ version of the holographic principle relates certain static DW's solutions of 3D gravity (Einstein or NMG) with scalar matter to the RG flows in specific deformed (supersymmetric) $CFT_2$ \cite{4}. These non-conformal $QFT_2$'s can be   realized as an appropriate perturbations of the ultraviolet (UV) $CFT_2$ by marginal or/and relevant operators $\hat{\Phi}_{\sigma}$ that break  2D conformal symmetry to the Poincare one :
\begin{eqnarray}
S_{pCFT_{2}}^{ren}(\sigma)=S_{CFT_2}^{UV}+\sigma(L_*)\int d^2x\sqrt{-g}\Phi_{\sigma}(x^i)\label{eq28}
\end{eqnarray}
The scale-radial duality \cite{VVB} allows to identify the ``running'' coupling constant $\sigma(L_*)$ of $pCFT_2$ (\ref{eq28})  with the scalar field $\sigma(z)$ and the RG scale $L_*$ with the scale factor $e^{\varphi(z)}$ as follows: $L_*=l_{pl}e^{-\varphi/2}$. 
This identification is based on the equivalence of the ``radial'' evolution equations (\ref{eq5}) and the Wilson RG equations for the $pCFT_2$:
\begin{eqnarray}
\frac{d\sigma}{dl}=-\beta(\sigma)=\frac{2\epsilon}{\kappa^2}\frac{W'(\sigma)}{W(\sigma)}\bigg(1-\frac{W^2(\sigma)\kappa^2}{2\epsilon m^2}\bigg),\quad\quad l=\ln L_*\label{eq30}
\end{eqnarray}
 It is evident that the zeros of the $\beta$-function (\ref{eq30}) $\sigma_{B}^*$ coincide with the NMG  vacuum of type ($a$) ($i.e.$ $W'(\sigma_{a}^*)=0$) or of the type ($b$) ($i.e.$ $W_{\pm}^2(\sigma_{\pm}^{*})=\frac{2\epsilon m^2}{\kappa^2}$). We also realize  that the anomalous dimensions $\Delta_{\Phi}$ of the operator $\Phi_{\sigma}(x^i)$ at each critical point:
\begin{eqnarray}
y(\sigma^*)=2-\Delta_{\Phi}(\sigma^*)=-\frac{d\beta(\sigma)}{d\sigma}\Big\vert_{\sigma=\sigma^*}\label{eq31}
\end{eqnarray}
are nothing but the parameters $\Delta_{\pm}^{a,b}(\sigma^*)$ and $y(\sigma_{a,b}^{*})$ given by eqs. (\ref{A5}), that determine the asymptotic behaviour at $z\rightarrow\pm\infty$ of  the matter 3D bulk gravity  field $\sigma(z)$.
  
  As it well known, when the explicit form of  the  $\beta(\sigma)$ - function  is given, say by eq. (\ref{eq30}), it provides the key ingredient that allows to further derive the free energy and certain thermodynamical characteristics of 2D \emph{classical} statistical model related (in its thermodynamic limit) to  the \emph{quantum} $pCFT_2$ in discussion\footnote{ indeed we have to consider the euclidean version of NMG such that the corresponding ``boundaries/horizons" of $(a)H_3$ are flat euclidean planes or spheres $S^2$.}. We are interested in the description of the scaling laws, critical exponents and the phase structure of particular $pCFT_2$  dual to NMG model (\ref{eq1}) with quadratic superpotential in the case  the range of its parameters $B$ and $D$ belongs to  the region (2). Following the standard RG methods (see for example \cite{muss}) we find that the singular part of the reduced free energy per 2D volume $F_s(\sigma)$ has the following simple form:
\begin{eqnarray}
F_s(\sigma)\approx (\sigma^2)^{\frac{1}{y_0}}\left((\sigma_+^*)^2-\sigma^2\right)^{\frac{2}{y_+}}\left(
|(\sigma_-^*)^2|+\sigma^2\right)^{\frac{2}{y_-}}\label{eq33}
\end{eqnarray} 
The critical exponents $\nu_A = \frac{1}{y_A}$  related to the correlation length singularities $\xi_A\approx(\sigma - \sigma_A^*)^{-\frac{1}{y_A}}$ at each critical point (i.e.the  NMG's vacua with $A=0,\pm $) are given by: 
\begin{eqnarray}
y_0=-\frac{4\epsilon B}{D\kappa^2\alpha_+\alpha_-},\quad\quad
y_+=\frac{8\epsilon B}{\alpha_+\kappa\sqrt{2\epsilon m^2}},\quad\quad 
y_-=-\frac{8\epsilon B}{\alpha_-\kappa\sqrt{2\epsilon m^2}}\label{eq34}
\end{eqnarray} 
For a particular choice of  the ``unitary" (for $\epsilon=-1 ,m^2<0 $)  DW's (\ref{eq10}) and of the coupling constant $\sigma$ within the range $0<\sigma<\sigma^*_+$ we have that $y_0<0$ (i.e. the IR $CFT_2$ with $\Phi_{\sigma}$ as irrelevant operator) and $0<y_+<2$ (i.e. the  UV $CFT_2$ with $\Phi_{\sigma}$ representing now a relevant operator). Therefore such DW describes massless RG flow from UV critical point $\sigma^*_+$ to the IR one $\sigma_0^*=0$.
An important characteristics of all the  massless flows  is the so called Zamolodchikov's central function :
\begin{eqnarray}
C(\sigma)=-\frac{3}{2G\kappa W(\sigma)}\bigg(1+\frac{\kappa^2W^2(\sigma)}{2\epsilon m^2}\bigg)\label{eq32}
\end{eqnarray}
which at the critical points $\sigma_{A_{\pm}}^*$ takes the values (\ref{eq28A}). It represents a natural generalization \cite{sinha} of the well known result for  $m^2\rightarrow\infty$ limit \cite{4},\cite{VVB}. According to its original 2D definition \cite{x} it is intrinsically related to the $\beta$-function:
\begin{eqnarray}
\beta(\sigma)=-\frac{4G \epsilon W(\sigma)}{3\kappa}\left(\frac{dC(\sigma)}{d\sigma}\right)
\end{eqnarray}
of the $pCFT_2$ dual of the NMG  model (\ref{eq1}). Taking into account the RG equations (\ref{eq30}) we  realize that :
\begin{eqnarray}
 \frac{dC(\sigma)}{dl} = -\frac{3}{4GW(\sigma)}\left(\frac{d \sigma}{dl}\right)^2 \label{eq29b}
\end{eqnarray}
and therefore when $W(\sigma)>0$ is positive (as in our example) the central function is decreasing during  massless flow we are discussing, i.e. we have $ c_+ > c_a$. 

Observe that for $ \sigma>\sigma^*_+$ and for $\sigma \rightarrow\infty$ the correlation length remains \emph{finite} due to the following ``resonance" property $\frac{1}{2y_0}+\frac{1}{y_+}+\frac{1}{y_-}=0$, specific for the quadratic superpotential we are studying. Hence  this region of the coupling constant space corresponds to the \emph{massive} phase of the $pCFT_2$, which is described ``holographically" by \emph{singular} DW  metrics giving rise to $(a)AdS_3$ space-time  with naked singularity, as one can see from the generic form (\ref{eq10}) of our DW's solutions. We have therefore  an example of phase transition from massive to massless phase that occurs at the UV critical point $\sigma_+^*$.  For the description of such phase transition we need \emph{two different} NMG solutions having coinciding boundary conditions $(\sigma_A^*,\Lambda_{eff}^A, \Delta_A)$ at their common boundary $z\rightarrow\infty$. 

We next briefly discuss the  possibility to extend  our $d=3$ superpotential constructions (\ref{eq4}) to the case of $d>3$ NMG models:
\begin{eqnarray}
S=\frac{1}{\kappa^2}\int d^dx\sqrt{-g}\Big\{R+\frac{1}{m^2}\left(R^{\mu\nu}R_{\mu\nu}-\frac{d}{4(d-1)}R^2\right)-\kappa^2\Big(\frac{1}{2}|\vec{\nabla}\sigma|^2+V(\sigma)\Big)\Big\}\label{dacao}
\end{eqnarray}
As in  $d=3$ case the static flat DW's solutions of such d-dimensional NMG model are defined by 
\begin{eqnarray}
ds^2=dz^2+e^{2\beta\varphi(z)}\eta_{ij}dx^idx^j,\quad \sigma=\sigma(z),\quad
\beta=\frac{1}{\sqrt{2(d-1)(d-2)}},\quad  \alpha=(d-1)\beta,\label{ddw}
\end{eqnarray}
that leads us to a system of \emph{second} order equations of the type (\ref{eq4}), but with different d-dependent coefficients. It is then natural to introduce the following generalization of $d=3$ NMG  superpotential and  of  the $I^{st}$ order system (\ref{eq5}) for arbitrary $d>3$ :  
\begin{eqnarray}
\dot{\varphi}&=&-2\kappa\alpha W(\sigma),\quad \dot{\sigma}=\frac{2}{\kappa}W'(\sigma)\left(1+\kappa^2\frac{(d-4)}{2(d-2)}\frac{W^2}{m^2}\right)\\
V(\sigma)&=&2(W')^2\left(1+\kappa^2\frac{(d-4)}{2(d-2)}\frac{W^2}{m^2}\right)^2-2\kappa^2\alpha^2W^2\left(1+\kappa^2\frac{(d-4)}{4(d-2)}\frac{W^2}{m^2}\right)\label{dsys}
\end{eqnarray}
which in $d=5$  case reproduces the Low-Zee  superpotential \cite{zee}  for the  Gauss-Bonnet(GB) extended 5D gravity. It is worthwhile to mention the well known fact (see ref. \cite{flat} for example) that for \emph{conformally flat} solutions (i.e. of vanishing $d>3$ Weyl tensor as in the case of the DW's (\ref{ddw})) the action of the Gauss-Bonnet-Einstein gravity  becomes identical to the $d>3$ NMG's one (\ref{dacao}). Therefore the solutions of the eqs.(\ref{dsys}) describe the flat static DW's of the both models. The form of the eqs.(\ref{dsys}) above makes evident that any given superpotential $W_d(\sigma)$ describe  qualitatively different matter potentials $V_d(\sigma)$ depending on the values of  $d=3,4,5$. Hence the properties of DW's solutions of corresponding NMG model's, as well as of the $\beta_d(\sigma)-$functions of their  $pCFT_d$ duals, are expected to be rather different depending on the space-time dimensions. We leave the problem of  the identification of these $QFT_d$ models and of the  geometrical NMG's description of their phase structure to our forthcoming paper \cite{nmgd}.    

Let us emphasize in conclusion  the advantages of the superpotential method  in  the study of the DW's properties of the NMG models (\ref{dacao})  as well as of the holographic RG flows in their dual $pCFT_d$ models. As we have shown on the example of 3D NMG model (\ref{eq1}) with quadratic $W(\sigma)$, the  DW's solutions  provide an important information about  the phase transitions in its dual 2D model. It is important to note  however that although we have recognized many of the  ingredients of the $AdS_3/CFT_2$  correspondence as central charges, scaling dimensions, free energy,etc. the answer to the question of whether  and under which conditions such correspondence takes place for the NMG model (\ref{eq1}) remains still open. The  complete identification and the  description of all the  properties  of the dual $pCFT_2$  in terms of the  NMG model's  solutions requires better understanding of the apparent ``unitarity discrepancy" that relates (1-loop) \emph{ unitary}  massive 3D gravity
to \emph{non-unitary} $CFT_2$'s of negative central charges  in the approximation of small  effective cosmological constants. Negative central charges are known to appear in different contexts in  the (supersymmetric) $CFT_2$'s. For example, the classical  and semi-classical  limits $\hbar\rightarrow 0 $ of  the central charges $c_q=1-6\frac{Q^2_{cl}}{\hbar}$  of the so called  minimal \emph{ unitary} Virasoro algebra models (as well as of their $N=1$ SUSY extensions) are big \emph {negative} numbers \cite{fat}. There exist also  families of non-unitary 2D models  representing interesting statistical mechanical problems, as for example the Lee-Yang ``edge singularity" $CFT_2$ of $c=-\frac{22}{5}$. We  remind these  facts  just to indicate few directions for further investigations  that  might result in the exact identification of the 2D QFT's duals  to 3D New Massive gravity models.

\emph{Acknowledgments} We are grateful to H.L.C. Louzada for his collaboration in the initial stage of this work and to C.P. Constantinidis for the discussions and for critical reading of the manuscript. This work has been partially supported by PRONEX project number 35885149/2006 from FAPES-CNPq (Brazil).   


\vspace{1 cm}
 


\begin{thebibliography}{99}
\bibitem{1} E.A. Bergshoeff, O. Hohm and P.K. Townsend.,Phys. Rev. Lett.\textbf{102}, 201301(2009);\\ 
Phy. Rev. \textbf{D79} 124042/2009
\bibitem{flat}E.A. Bergshoeff, O. Hohm and P.K. Townsend,Gravitons in Flatland, arXiv:1007.4561.
\bibitem{des}S.Deser,Phys. Rev. Lett.\textbf{103}, 101302(2009).
\bibitem{oda}I.Oda, JHEP{0905},064 (2009).
\bibitem{2}G.Clement,Class.Quant.Grav.\textbf{26},105015(2009),Phys.Rev.Lett.\textbf{102},201301(2009); 
E.Ayon-Beato,G.Giribet,M.Hassaine,JHEP{0905},029 (2009);J.Oliva,D.Tempo and R.Troncoso,JHEP 0907:011(20009);H.Ahmedov and A.N.Aliev,The general type N solutions of New Massive Gravity,arxiv:1008.0303.
\bibitem{3}H.L.C. Louzada, U. Camara dS and G.M. Sotkov, Phys. Lett. \textbf{686 B} (2010) 268.
\bibitem{4}D.Z. Freedman, S.S. Gubser, K.Pilch and N.P. Warner, Adv. Theor. Math. Phys. 3: 363-417(1999). \bibitem{malda}J.Maldacena, Adv.Theor.Math.Phys.2:231-252(1998).
\bibitem{witt}E.Witten, Adv.Theor.Math.Phys.2:253(1998).
\bibitem{pol}S.S.Gubser,I.R.Klebanov and A.M.Polyakov, Phys.Lett.\textbf{428 B}(1998) 105.
\bibitem {zee} I.Low and A.Zee, Naked Singularity and Gauss-Bonnet term in Brane world scenarios, NPB$\{585\}\{2000\}\{395-401\}$.
\bibitem{5}D.Z.Friedman,C.Nunez,M.Schnabl,K.Skenderis,Phys.Rev.\textbf{D69},104 027(2004);\\ M.Cvetic,S.Griffies,S.J.Rey, Nucl. Phys  \textbf{B381}{1992}{301} 
\bibitem{6}M. Cvetic and H.H. Soleng, Phys. Rep. 282(1997) 159 and references therein.
\bibitem{BF}P.Breitenlohner and D.Z.Freedman,PLB$\{115\}\{1982\}\{197\}$;\\
Ann.Phys.144(1982)249.
\bibitem{7}J.D.Brown and J.W. York, Phys. Rev.\textbf{D47}, 1407 (1993).
\bibitem{bala}V.Balasubramanian and P.Kraus, Commun. Math. Phys. \textbf{208} (1999) 413.
\bibitem{8}O. Hohm and A. Tonni, JHEP  1004:093(2010) .
\bibitem{town}K.Skenderis and P.K.Townsend,Gravitational stability and Renormalization group flow,PLB$\{468\}\{1999\}\{46\}$
\bibitem{9}J.D.Brown and M.Henneaux, Commun.Math.Phys.\textbf{104},207 (1986).
\bibitem{VVB} J. de Boer, Fortsch. Phys. 49:339-358(2001), hep-th/0101026;\\
  E. Verlinde, H. Verlinde and J. de Boer , JHEP  0008:003(2000) ,hep-th/9912012.
\bibitem{x} A.B. Zamolodchikov, Sov. Phys. JETP Lett. {\bf 43}(1986)1731;\\
Sov.J.Nucl.Phys.\textbf{46}(1987),1090.
\bibitem{muss} G.Mussardo, Statistical Field Theory, Oxford University Press Inc., New York, 2010.
\bibitem{sinha} A.Sinha, JHEP  1006:061(2010). 
\bibitem{fat}V.Fateev and S.Lukyanov, Sov.Sci.Rev.A.Phys.Vol.15 (1990),pp.1-117(see pg.18);\\
         G.Sotkov and M.Stanishkov, NPB {\bf 356},(1991),439.
\bibitem{nmgd}U.Camara dS, C.P.Constantinidis and G.M.Sotkov, Domain Walls and Holographic RG flows in D-dimensional New Massive Gravity(in preparation).
\end{thebibliography}
\end{document}